\documentclass[sn-nature]{sn-jnl}

\usepackage{graphicx}
\usepackage{multirow}
\usepackage{amsmath,amssymb,amsfonts}
\usepackage{amsthm}
\usepackage{mathrsfs}
\usepackage[title]{appendix}
\usepackage{xcolor}
\usepackage{textcomp}
\usepackage{manyfoot}
\usepackage{booktabs}
\usepackage{algorithm}
\usepackage{algorithmicx}
\usepackage{algpseudocode}
\usepackage{listings}

\theoremstyle{thmstyleone}

\theoremstyle{thmstyletwo}

\theoremstyle{thmstylethree}

\begin{document}

\title[Article Title]{Phononic materials with effectively scale-separated hierarchical features using interpretable machine learning}

\author[1]{\fnm{Mary V.} \sur{Bastawrous}}
\equalcont{These authors contributed equally to this work.}

\author[2]{\fnm{Zhi} \sur{Chen}}
\equalcont{These authors contributed equally to this work.}

\author[3]{\fnm{Alexander C.} \sur{Ogren}}

\author[3]{\fnm{Chiara} \sur{Daraio}}

\author[2]{\fnm{Cynthia} \sur{Rudin}}

\author*[1]{\fnm{L. Catherine} \sur{Brinson}}\email{cate.brinson@duke.edu}

\affil*[1]{\orgdiv{Department of Mechanical Engineering and Materials Science}, \orgname{Duke University}, \orgaddress{\street{305 Teer Engineering Building}, \city{City}, \postcode{NC 27708}, \state{North Carolina}, \country{USA}}}

\affil[2]{\orgdiv{Department of Computer Science}, \orgname{Duke University}, \orgaddress{\street{308 Research Dr}, \city{Durham}, \postcode{NC 27705}, \state{North Carolina}, \country{USA}}}

\affil[3]{\orgdiv{Division of Engineering and Applied Science}, \orgname{California Institute of Technology}, \orgaddress{\street{1200 E. California Boulevard}, \city{Pasadena}, \postcode{ CA 91125}, \state{California}, \country{USA}}}

\abstract{
Manipulating the dispersive characteristics of vibrational waves is beneficial for many applications, e.g., high-precision instruments. architected hierarchical phononic materials have sparked promise tunability of elastodynamic waves and vibrations over multiple frequency ranges. In this article, hierarchical unit-cells are obtained, where features at each length scale result in a band gap within a targeted frequency range. Our novel approach, the ``hierarchical unit-cell template method,'' is an interpretable machine-learning approach that uncovers global unit-cell shape/topology patterns corresponding to predefined band-gap objectives. A scale-separation effect is observed where the coarse-scale band-gap objective is mostly unaffected by the fine-scale features despite the closeness of their length scales, thus enabling an efficient hierarchical algorithm. Moreover, the hierarchical patterns revealed are not predefined or self-similar hierarchies as common in current hierarchical phononic materials. Thus, our approach offers a flexible and efficient method for the exploration of new regions in the hierarchical design space, extracting minimal effective patterns for inverse design in applications targeting multiple frequency ranges.
}


\maketitle

\section{Introduction}\label{sec1}

\begin{figure}[htbp!]
\centering
\includegraphics[scale=0.6,trim={0cm 0cm 0cm 0cm},clip]{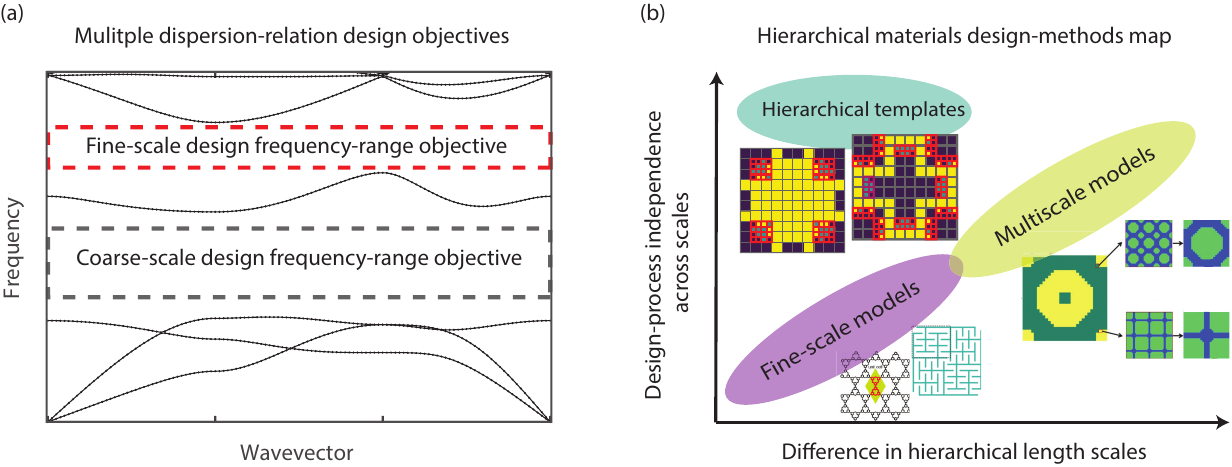}

    \caption{\textbf{(a)} A schematic for a unit-cell dispersion relation indicating multiband design objectives for coarse-scale and fine-scale patterns. \textbf{(b)} A design-space map for hierarchical materials with different classes indicated (included insets from left to right are from refs. \cite{mousanezhad2015honeycomb, liu2018fractal, liang2020design}, respectively). The horizontal axis represents the extent of difference in hierarchical length scales (from close hierarchical scales within the same order of magnitude on the left to significantly different hierarchical scales varying by several orders of magnitude on the right). The vertical axis represents the extent of independence of design processes at each hierarchical scale from the remaining scales.  Models with fine-scale grids are needed to analyze hierarchical materials with close scales, which means that design processes for the hierarchical scales are interdependent (bottom-left corner). Conversely, hierarchical materials with significantly different scales are usually analyzed with multi-scale models and the design process at each hierarchical length scale can be performed relatively independently from the remaining scales (top-right corner). Our developed approach, ``hierarchical templates,'' fits in the top-left corner of the design-space map as it represents handling hierarchical materials with close scales but, as we show, the design process for different scales can still be carried out independently (consecutively from coarse to fine).}
    \label{fig:Fig_01}
\end{figure}

Phononic materials are architected materials that are used to manipulate the dispersive properties of vibrational waves, i.e., phonons \cite{Deymier2013-cb, Hussein2014-ac}. The tunability of the phononic band diagrams of these materials may lead to interesting properties such as bandgaps (frequency bands where waves attenuate), negative refraction, and wave localization.
Typically, phononic materials are built by periodically tessellating a unit-cell. The wave-dispersion properties of the periodically repeating unit-cell can be obtained by performing Bloch unit-cell analysis \cite{Deymier2013-cb}. The material distribution and geometry of the unit-cell can be tailored to tune these wave-dispersion characteristics at desired wavelength or frequency ranges. 
The spatial scale of features in the unit-cell determine the wavelengths (and therefore frequencies) of the waves they interact with.
Hierarchical phononic materials \cite{footnote1} exploit this relationship to provide multi-band wave dispersion tunability across different wavelength/frequency ranges by engineering multiscale patterns into the unit-cell \cite{Xu2014-gb,Mousanezhad2015-ib,Xu2019-hq,Liang2019-kz,Liang2020-as,Zhang2020-tz,Sun2020-qd,Zhang2020-tz,Lee2020-hp,Mei2021-gr,Sun2022-sl} [see \textbf{\autoref{fig:Fig_01}(a)}].

Hierarchical phononic materials maybe such that the length scales are orders of magnitude apart \cite{Liang2019-kz,Liang2020-as,Mei2021-gr,Zhang2020-tz}. In this case,  the scale-separation principle is in effect and the frequency ranges where each scale is influential are well separated. The two length scales may then be designed independently from each other. 
In other cases, the hierarchical length scales are relatively close (within an order of magnitude) \cite{Xu2014-gb,Mousanezhad2015-ib,Xu2019-hq,Zhang2020-tz,Sun2020-qd,Zhang2020-tz,Lee2020-hp,Mei2021-gr,Sun2022-sl}, as illustrated in \textbf{\autoref{fig:Fig_01}(b)} and thus may strongly influence overlapping frequency ranges \cite{Xu2014-gb,Mousanezhad2015-ib,Lee2020-hp,Sun2022-sl}. 
This type is generally easier to fabricate.
The design process is then typically performed for multiple length scales simultaneously on fine discretizations that resolve the finest utilized length scale. This incurs significant computational cost and prohibits exploration of the entire design space [\textbf{\autoref{fig:Fig_01}(b)}]. This challenge has limited the advancement of hierarchical phononic materials in the following ways: 1) Current results have so far been limited to hierarchical unit-cells with predetermined hierarchical patterns, e.g., self-similar and fractal patterns \cite{Mousanezhad2015-ib,Liu2018-gc,Man2018-dc,Man2021-hm}; 2) There is no knowledge of patterns in the broader hierarchical design space which would satisfy multiband design objectives, e.g., bandgaps at different frequency ranges; 3) The design process has been dominantly trial-and-error driven and highly reliant on designers' experience. 
 
Machine learning methods have recently emerged as an attractive alternative to design phononic materials \cite{He2021-va,Zhang2021-rf,Chen2021-zx,Jin2022-kp,finolDeepConvolutionalNeural2019a}, and more generally architected materials \cite{Guo2020-ul,Shin2021-bu,PhysRevLettMastrigt,jiaoMechanicalMetamaterials2023}. One of the advantages of machine-learning algorithms is that they may have the ability to map the design space, learn the underlying rules for successful designs, and possibly generate new ones, thus generating new knowledge. However, machine learning methods used for phononic materials have so far been employed as black-box models that either map input feature sets to output properties in case of the forward property-prediction problem, or output designs to input desirable design objectives in case of the inverse design problem, without much insight into the learned rules \cite{He2021-va,Zhang2021-rf,Jin2022-kp}. \textit{Interpretable} machine learning methods, on the other hand, use custom constraints to incorporate materials and engineering domain expertise which allow users to understand the rationale and patterns for successful designs. See \cite{rudin2022interpretable} for an overview of interpretable machine learning methods. 

The unit-cell template method \cite{Chen2021-zx} was specifically developed as an interpretable machine learning method for designing dual-phase phononic materials. 
The method uncovers global patterns of constituent materials in dual-phase pixelated phononic-material unit-cells to meet a given design objective.
To satisfy the objective, the templates identify \textit{fixed regions} where specific materials are required, and \textit{free regions} where a choice of material is allowed. 
By varying the material in these free regions, multiple unit-cells with the same desired bandgap can be generated. 
The unit-cell template method is described in further detail in \cite{Chen2021-zx}, along with the resulting optimal template sets for dual-phase phononic-materials with different targeted bandgap ranges. 
Notably, the method demonstrates robustness to finer features in the free-pixel regions, enabling its extension (in this work) to discover and design hierarchical phononic material unit-cells.

\begin{figure}[htbp!]
\centering
\includegraphics[scale=0.6,trim={0cm 0cm 0cm 0cm},clip]{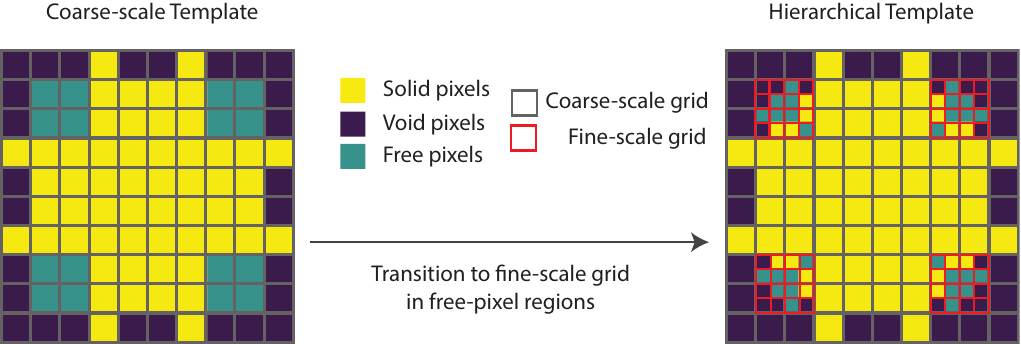}
    \caption{Outline for the hierarchical unit-cell template method: On the left, a coarse-scale template with fixed pixels as well as free pixels is illustrated. Note that the free (green) pixels can be assigned the properties of a solid material or void and the target coarse-scale bandgap objective would still be achieved. As a reminder, the coarse unit-cell template was designed to satisfy a coarse-scale design objective, e.g., a specific bandgap exhibited by the periodic wave-propagating medium. Upon further refining the green free-pixel regions, fine-scale templates are obtained by our method to satisfy an additional design objective, e.g., another bandgap in a different frequency range. Thus, the hierarchical template, including both coarse and fine-scale grids, satisfies both design objectives.}
    \label{fig:Fig_02}
\end{figure}

Here, we develop a novel approach to obtain hierarchical phononic material unit-cells, called \textit{hierarchical phononic-material templates}, which targets bandgaps in two different frequency regimes. In this new method, the unit-cell templates reported in \cite{Chen2021-zx} are modified to have multiple-scale grids that delineate coarse-scale and fine-scale regions. The fine-scale regions are designed into the coarse-scale free-pixel design areas. See \textbf{\autoref{fig:Fig_02}} for a schematic illustration of this approach. 
The design objective here is to achieve predefined bandgaps in different frequency ranges where each hierarchical scale is designed towards satisfying a single bandgap. Thus, the hierarchical unit-cell templates enable us to discover  the effective patterns at multiple scales.   

Our approach, \textit{hierarchical phononic-material templates}, offers several important benefits over past approaches:
\begin{itemize}
    \item It allows us to \textbf{search vast areas of the hierarchical design space of phononic materials that have not been previously explored}. This is because our search space is much larger. We do not use predetermined or repetitive patterns, nor do we rely on designers' expertise and/or trial-and-error \cite{mousanezhad2015honeycomb, liu2018fractal, liang2020design}.
    \item Our \textbf{multiscale} technique \textbf{reduces computational time}, because each scale is handled independently, even though the scales are close. The scales are designed coarse-to-fine. The coarse scale is designed to preserve a specific bandgap regardless of what will be assigned in the future at any finer scale. That is, the coarse scale design objective is \textbf{robust} to any smaller pixels assigned at any finer scale.
    \item  Therefore, there is an \textbf{effective scale separation consistently and quantitatively demonstrated}, which makes our approach different from existing works \cite{Sepehri2020-np,Miniaci2021-hk}, where the coarse-scale objective preservation is investigated a on a case-by-case basis for each design. This scale-separation effect is at work between the hierarchical length scales, despite the fact that the length scales in question are strongly interacting with overlapping frequency ranges\textemdash the minimum feature size in the fine scale is only half the minimum feature size in the coarse length scale.
     
\end{itemize}
To help put our method in perspective, \autoref{fig:Fig_01}(b) shows where the hierarchical templates fit in the map of design methods for hierarchical phononic materials. It shows that although the length scales are close and have been typically investigated using fine grids for the design and search process, our approach handles each scale independently and consecutively. Thus, we present a new design paradigm that occupies a previously vacant spot in the space of design methods for hierarchical phononic materials, and brings huge computational savings because the exponentially larger design space at the finer scale is efficiently characterized by the hierarchical templates.

\section{Results}\label{sec2}

\subsection{Parent coarse-scale templates}
\begin{figure}[htbp!]
\centering
\includegraphics[scale=0.5,trim={0cm 0cm 0cm 0cm},clip]{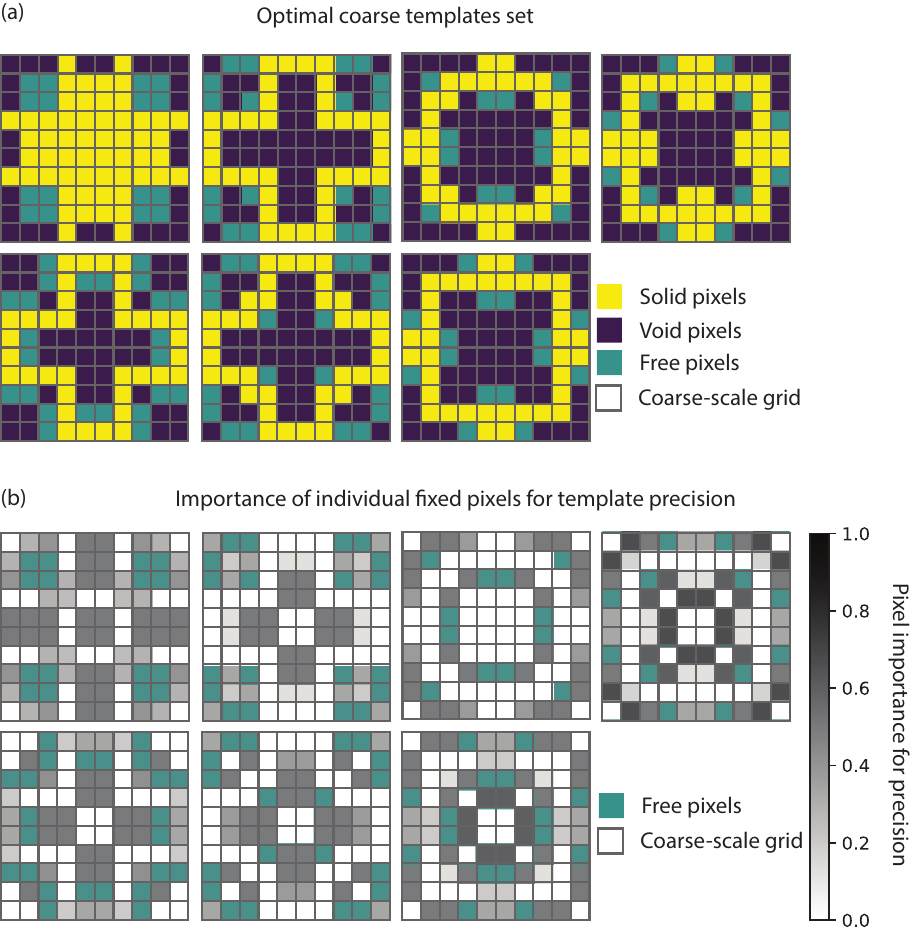}
    \caption{\textbf{(a)} Optimal set of coarse-scale templates obtained such that the design objective of having a single bandgap on a minimum of $60\%$ of the normalized-frequency interval of [0.28, 0.42]. \textbf{(b)} Influence of perturbing each fixed pixel on the templates precision, with darker pixels indicating higher importance.}
    \label{fig:Fig_04}
\end{figure}

We start by finding the parent templates that will satisfy the coarse-scale bandgap objectives in \autoref{tab:TempOptParams}. The resulting optimal template set is shown in \autoref{fig:Fig_04} (a). The parent templates' precision is $100\%$ [shown in \autoref{tab:CoarseTempRes_short}], which means that all the coarse-scale unit-cells generated from these templates meet the coarse-scale bandgap design objective. In the templates, the yellow color indicates solid pixels, while the purple color indicates void pixels. The green color indicate pixels which maybe occupied by either phase. The unit-cells that can be generated from the optimal templates set jointly by varying the green free-design pixels cover 28 unit-cells out of a total of 60 unit-cells in the design space, ($\sim50\%$ of the total number of unit-cells that satisfy the specified bandgap objective). Analyzing the templates, the fixed pixels seems to indicate an importance of cross-like features, where the crosses could be hollow or filled, and could have jagged or simple right-angled corners. 
\begin{table}[htbp!]
\centering
\begin{tabular}{p{2.9cm}|p{2.9cm}|c|c|p{2.9cm}}
\hline
Target normalized-frequency ($f L c^{-1}$) range   & Target coverage of target range (\%) & Precision & Support & Total \# valid designs in the design space\\ \hline 
{[}0.28, 0.42{]} & 60\% & 100.0\% & 28   & 60\\   \hline 
\end{tabular}
\caption{Results of parent coarse-scale template optimization. Coverage indicates how much of the target frequency range is covered by the bandgap. The template-set support indicates how many unit-cells can be generated from the template set.}
\label{tab:CoarseTempRes_short}
\end{table}

The unit-cell templates indicate rules for assigning material constituents to different pixels in the unit-cells, as they determine whether a solid or void phase must exist at these locations, or whether they can be freely designed to be either. To further understand the role of each pixel in the templates, we perform an experiment where each fixed pixel in the templates is perturbed (switched from solid to void, or vice versa). The resulting precision of the perturbed template is shown on the gray-scale colorbar in \autoref{fig:Fig_04} (b). The darker the pixel color, the lower the precision of the perturbed template, and therefore the more the contribution of this fixed pixel to the overall template precision. Some pixels are colored in white, i.e., they have no contribution to the template precision.
It may seem contradictory that these pixels are assigned to be fixed pixels by the template algorithm that seeks to maximize the template support (number of green pixels) while maintaining the minimum precision. However, upon closer inspection, it is found that these white pixels preserve the connectivity and minimum feature-size constraints of the template. Therefore, each fixed pixel in the template either contributes to the overall template precision (some pixels more than others), or has a role in maintaining the connectivity and feature size constraints. In addition to the enhanced understanding of individual pixel contributions to the bandgap objective, this analysis can also highlight the most effective routes for opening or closing the bandgaps. This is because the individual pixels that cause the highest change in the precision would be the most promising ones to perturb in case it is desirable to switch on/off the associated bandgaps.

\begin{figure}[htbp!]
\centering
\includegraphics[scale=0.45,trim={0cm 0cm 0cm 0cm},clip]{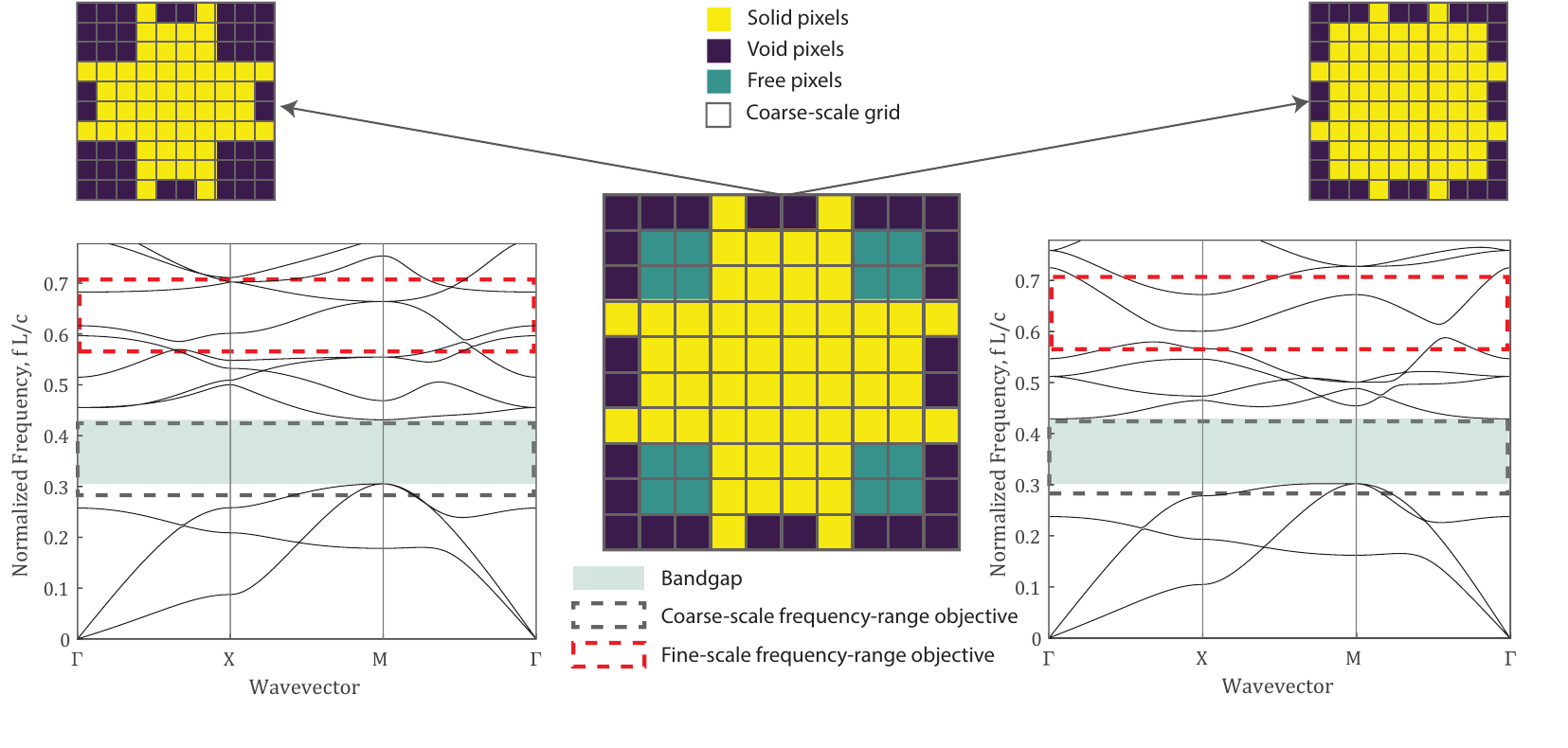}

    \caption{First example of a coarse-scale template obtained such that the design objective of having a single bandgap on a minimum of $60\%$ of the normalized-frequency interval of [0.28, 0.42] (gray region) is satisfied. Two unit-cells from this template, representing minimum and maximum solid fraction, along with their respective dispersion curves are shown. A second bandgap objective (highlighted by the red dashed region) is later defined for the hierarchical template derived from this coarse template (see \autoref{fig:Fig_03}) and is shown to be unmet at the coarse scale.}
    \label{fig:Fig_05}
\end{figure}

\begin{figure}[htbp!]
\centering
\includegraphics[scale=0.45,trim={0cm 0cm 0cm 0cm},clip]{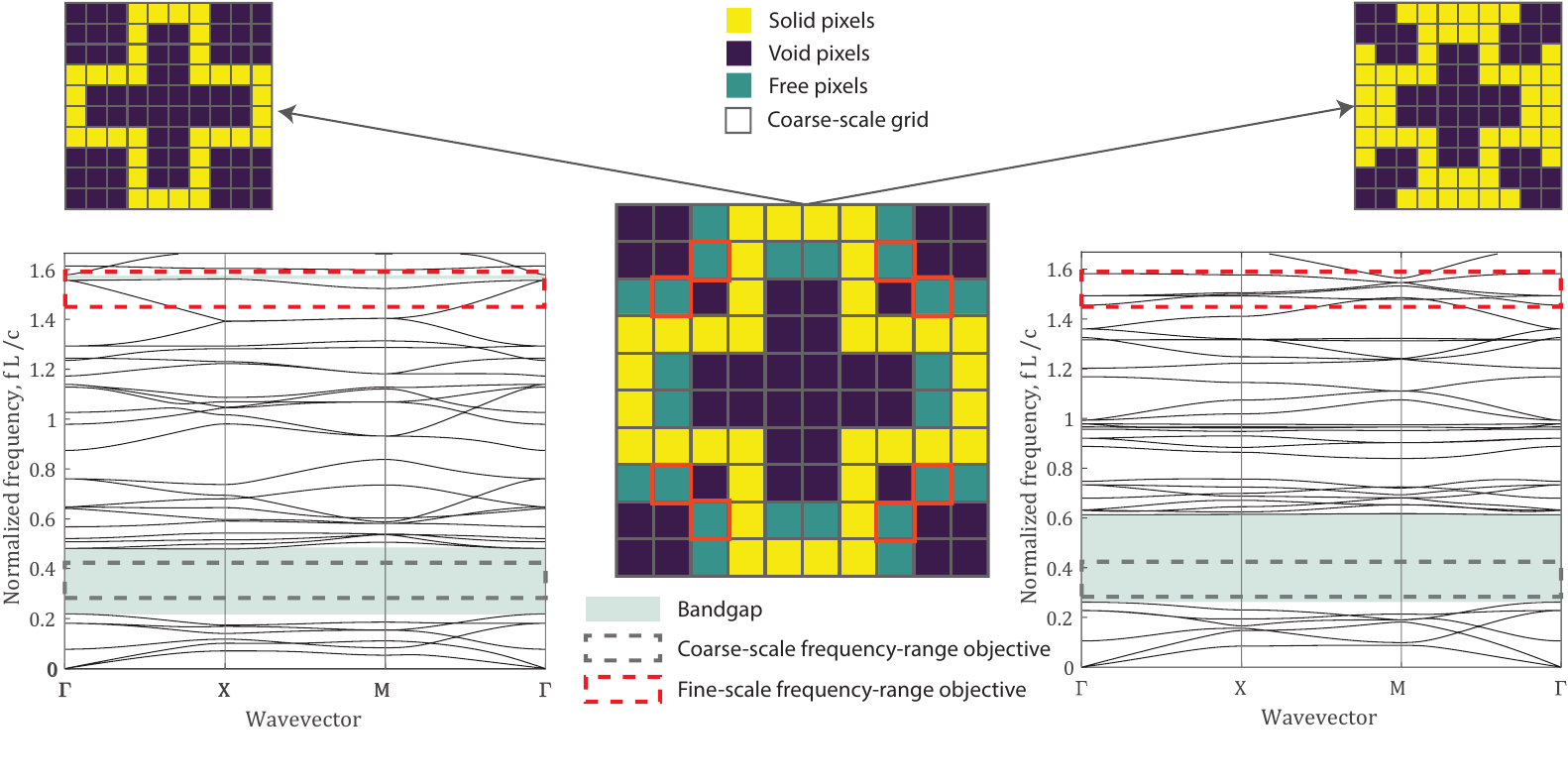}

    \caption{Second example of a coarse-scale template obtained such that the design objective of having a single bandgap on a minimum of $60\%$ of the normalized-frequency interval of [0.28, 0.42] (gray region) is satisfied. Two unit-cells generated from this template, representing minimum and maximum solid fraction, along with their respective dispersion curves are shown. A second bandgap objective (highlighted by the red region) is later defined for the hierarchical template derived from this coarse template (see \autoref{fig:Fig_04}) and is shown to be unmet at the coarse scale. Note that the free pixels highlighted with bright red outlines cannot be flipped to solid as it would lead to zero-dimensional points of connection at the corners.}
    \label{fig:Fig_06}
\end{figure}

We next examine two parent coarse-scale templates from \autoref{fig:Fig_04} (a) (shown in \autoref{fig:Fig_05} and \autoref{fig:Fig_06}). 
Two unit-cells per template are generated by assigning the free pixels as entirely void or solid, and their band diagrams computed, as shown in \autoref{fig:Fig_05} and \autoref{fig:Fig_06}.
The gray and red dashed rectangles illustrate target frequency ranges for coarse and fine scales (for use in the hierarchical template). Bandgaps, highlighted in green, demonstrate the fulfillment of the coarse-scale design objective, covering at least 60\% of the targeted frequency range.
Below the bandgap, the dispersion curves for each template's corresponding unit-cells are largely similar.
Note that the coarse-scale templates do not satisfy the fine-scale objective (in fact, we have chosen a design objective that cannot be satisfied by any coarse-scale design).
These will be discussed further in the next section.

These examples also demonstrate additional design flexibility, such as material density.
In \autoref{fig:Fig_05}, the left-side unit-cell has a filling density of $56\%$, and the right-side unit-cell has a filling density of $72\%$, both achieving the same bandgap objective. 
Note that for the second chosen template in \autoref{fig:Fig_06}, filling in the green pixels surrounded by the red squares would result in zero-dimensional features that introduce modeling and fabrication issues.
Therefore, such unit-cells are avoided and are not counted within the template's support.

\subsection{Hierarchical multi-scale templates}
In this section, we demonstrate the ability to design for an additional bandgap at a higher frequency using hierarchical phononic material templates. 
Following the previously described algorithm and problem setup discussed in Sections \ref{sec:method} and \ref{sec:setup} and highlighted in \autoref{fig:Fig_03}, we obtain the optimal hierarchical template set for our target frequency ranges, minimum bandgap width, and template precision as shown in \autoref{tab:TempOptParams}.
The precision and support of the resulting optimal template sets are shown in \autoref{tab:HierTempRes}. 
The optimal set support ranges from close to $50\%$ to $100\%$ of the design space.
All unit cells covered by the templates successfully meet both bandgap design objectives, (indicated by 100\% precision).
This remarkably demonstrates an \textit{effective scale separation} between the fine-scale and coarse-scale patterns in the hierarchical template sets, where the fine-scale patterns do not compromise the design objectives achieved by the coarse-scale patterns.

\begin{table}[htbp!]
\centering
\begin{tabular}{p{1.4cm}|p{2.5cm}|p{2.2cm}|c|c|p{2.2cm}}
\hline
Hierarchical template \# &Target normalized-frequency $f L c^{-1}$ range   & Target coverage of target range (\%)  & Precision & Support & Total \# valid designs in the design space  \\\hline
1& {[}0.28, 0.42{]} & 60\% & 100.0\%  & 28     & 60\\ 
& {[}0.57, 0.71{]} & 35\% & 100.0\%  &  43      & 44\\ 
2 & {[}0.28, 0.42{]} & 60\% & 100.0\%  & 28      & 60\\ 
& {[}1.45, 1.59{]}  & 35\%  & 100.0\%  & 75      & 147                                                                        \\ \hline
\end{tabular}
\caption{Hierarchical-template optimization results; Coverage indicates how much of the target frequency range is covered by the bandgap. The template-set support indicates how many unit-cells can be generated from the template set.}
\label{tab:HierTempRes}
\end{table}

The fact that the length scale for the larger grid is only twice that of the smaller grid in which the fine-scale features are defined makes this case  unlike traditional scale separation where the length scales are vastly different from each other and thus are naturally independent. It should also be emphasized that this effect exists independently, without the use of other aiding factors, e.g., viscoelastic damping, to preserve the coarse-scale bandgaps. 

To understand the scale-separation effect, its applicability, and limitations in our hierarchical templates, we performed a numerical experiment where we investigated the robustness of the coarse-scale bandgap objective to the introduction of fine-scale features into the free-pixel regions in the coarse-scale parent templates. This is performed by refining the grid in the free-pixel regions, generating all possible fine-scale unit-cells from the refined free-pixel regions, and evaluating the coarse-scale bandgap objective precision in the generated fine-scale unit-cells. 
This procedure is repeated for different values of the target minimum bandgap width defined as coverage percentage of the bandgap target range. The results of this numerical study are shown in \autoref{tab:CoarseTempRes}. The \textit{fine-scale transfer precision} is defined as the probability that the fine-scale unit-cells generated from the refined free-pixel regions in the coarse-scale template would satisfy the coarse-scale bandgap objective. Thus, the fine-scale transfer precision gives us a direct measure of the robustness of the coarse-scale templates precision to the introduction of fine-scale features not included in the training dataset. It is seen that the fine-scale transfer precision depends on the coarse-scale objective bandgap minimum width. The precision values are as high as $90\%$ if the bandgap covers $75\%$ of the target coarse-scale range. In our chosen case where the bandgap covers $60\%$ of the target range, the fine-scale transfer precision drops by only $5\%$. If we decrease the bandgap width by more than half of the initial width, the fine-scale transfer precision is still as high as $63\%$. 
Thus, this study offers valuable insight into the robustness of the scale-separation objectives and how fast the method is expected to break down. 

\begin{table}[htbp!]
\centering
\begin{tabular}{p{2.3cm}|p{2cm}|p{1.5cm}|c|p{2.3cm}|p{1.5cm}}
\hline
Target normalized-frequency ($f L c^{-1}$) range   & Target coverage of target range (\%) & Coarse-scale precision & Support & Total \# valid designs in the design space & Fine-scale transfer precision\\ \hline
{[}0.28, 0.42{]} & 75\% & 100.0\% &  22  & 46 & 90.0\\         \hline                                
{[}0.28, 0.42{]} & 60\% & 100.0\% & 28   & 60 & 85.1\\         \hline 
{[}0.28, 0.42{]} & 50\% & 100.0\% &  48  & 86 & 78.7\\         \hline  
{[}0.28, 0.42{]} & 35\% & 100.0\% & 82  & 140 & 63.0\\                
\end{tabular}
\caption{Results of parent coarse-scale template optimization when bandgap width changes. Fine-scale transfer precision, defined as the probability that fine-scale unit-cells generated from a coarse-scale template satisfy that coarse template's bandgap objective, is also computed. The last column shows how the scale-transfer precision increases by increasing the coarse-scale bandgap width.}
\label{tab:CoarseTempRes}
\end{table}

There are some important practical advantages of the scale-separation effect, as discussed earlier: computation is exponentially easier if the fine-scale computations are done only after the coarse scale is complete. It also allows designers to explore hierarchical designs spaces that do not necessarily exhibit repetitive or self-similar hierarchical patterns at multiple scales. Our method also illuminates underlying patterns in the unit-cells for specific design objectives.

\begin{figure}[htbp!]
\centering
\includegraphics[scale=0.45,trim={0cm 0cm 0cm 0cm},clip]{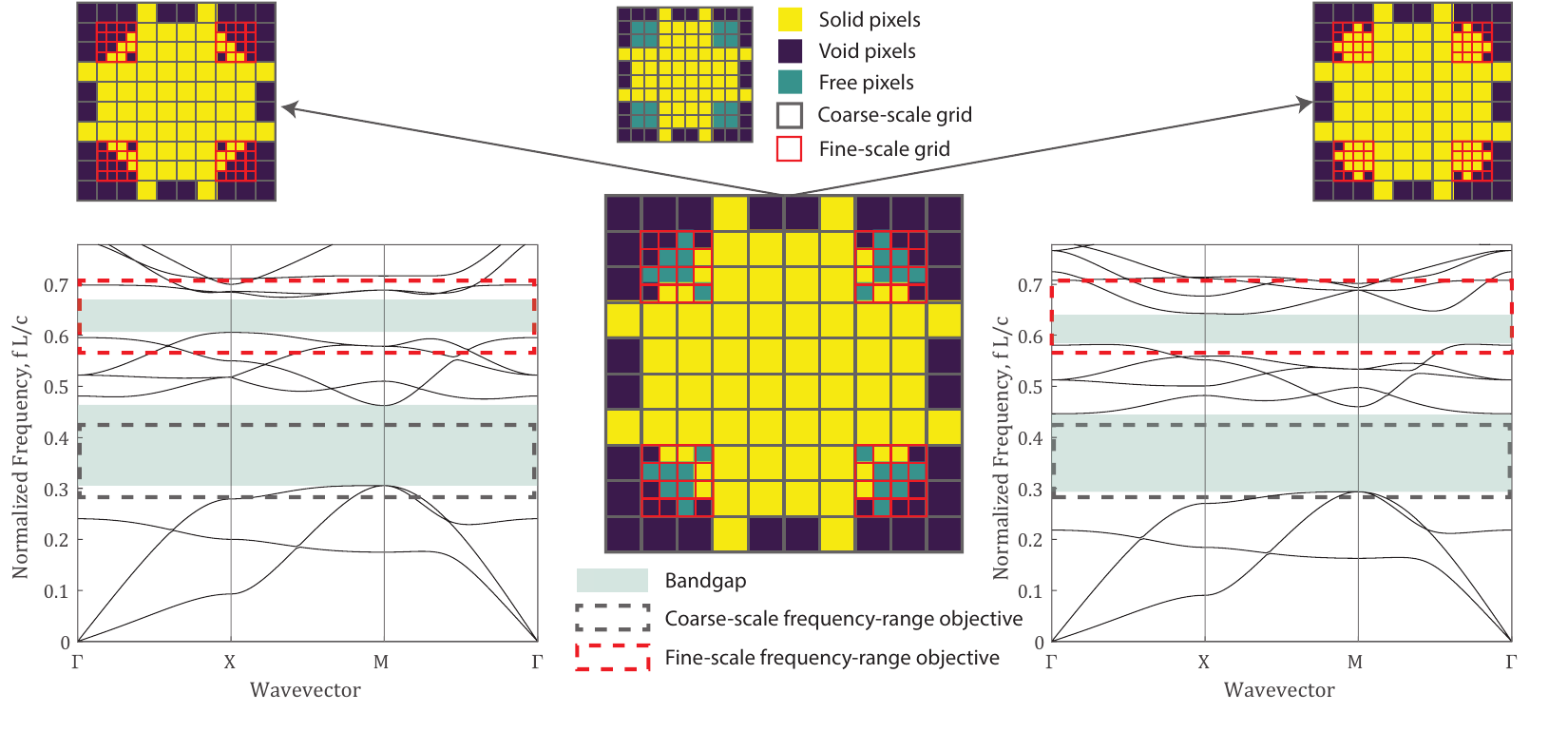}
    \caption{First example of a hierarchical template obtained from the parent coarse-scale template in \autoref{fig:Fig_05} such that the design objective of having a bandgap on a minimum of $60\%$ of the normalized-frequency interval of [0.28, 0.42] and a bandgap on a minimum of $35\%$ of the normalized-frequency interval [0.5657, 0.7071] is satisfied. Two unit-cells generated from this template along with their respective dispersion curves are shown. The coarse-scale bandgap objective (gray region) as well as the fine-scale bandgap objective (red region) are met.}
    \label{fig:Fig_07}
\end{figure}

\begin{figure}[htbp!]
\centering
\includegraphics[scale=0.45,trim={0cm 0cm 0cm 0cm},clip]{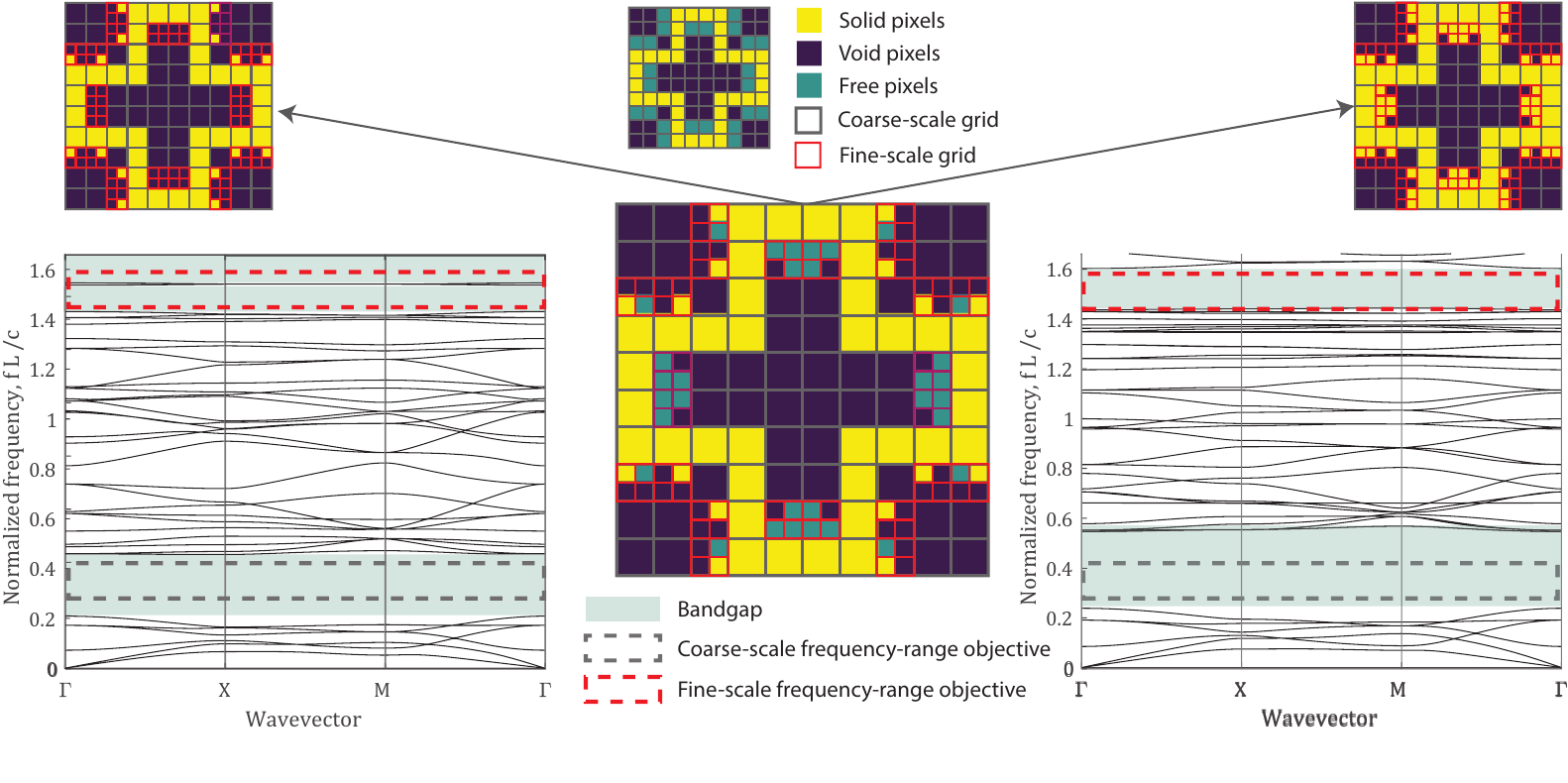}

    \caption{Second example of a hierarchical template obtained from the parent coarse template in \autoref{fig:Fig_06} such that the design objective of having a bandgap on a minimum of $60\%$ of the normalized-frequency interval of [0.28, 0.42] and a bandgap on a minimum of $35\%$ of the normalized-frequency interval [1.4496, 1.591] are satisfied. Two unit-cells generated from this template along with their respective dispersion curves are shown. The coarse-scale bandgap objective (gray region) as well as the fine-scale bandgap objective (red region) are met.}
    \label{fig:Fig_08}
\end{figure}

Two examples of the hierarchical templates that satisfy the predefined design objectives for both the coarse and fine-scale grids are shown in \autoref{fig:Fig_07} and \autoref{fig:Fig_08}. The corresponding parent coarse templates are also illustrated in these figures (top middle portion). For clarity, the fine-scale grid is illustrated by a red color, while a gray grid color is reserved for the coarse-scale grid. As in the coarse-scale demonstration, again for each hierarchical  template in both \autoref{fig:Fig_07} and \autoref{fig:Fig_08}, two unit-cells are generated by assigning all free fine-scale pixel regions to be either void or solid. The dispersion relations for each unit-cell are also shown in \autoref{fig:Fig_09} and \autoref{fig:Fig_10}. It can be observed that all unit-cells satisfy both the coarse-scale and fine-scale design objectives. Note that the coarse-scale unit-cells in \autoref{fig:Fig_05} and \autoref{fig:Fig_06} clearly demonstrated that the higher frequency bandgap was not achieved for either case having the coarse-scale pattern alone. Also critically, the insertion of the fine-scale grid and the opening of the higher frequency bandgap shown in \autoref{fig:Fig_07} and \autoref{fig:Fig_08} does not disrupt the existing lower-frequency bandgap from the coarse-scale grid.

\section{Numerical and experimental validation}

\begin{figure}[htbp!]
\centering
\includegraphics[scale=0.6,trim={0cm 0cm 0cm 0cm},clip]{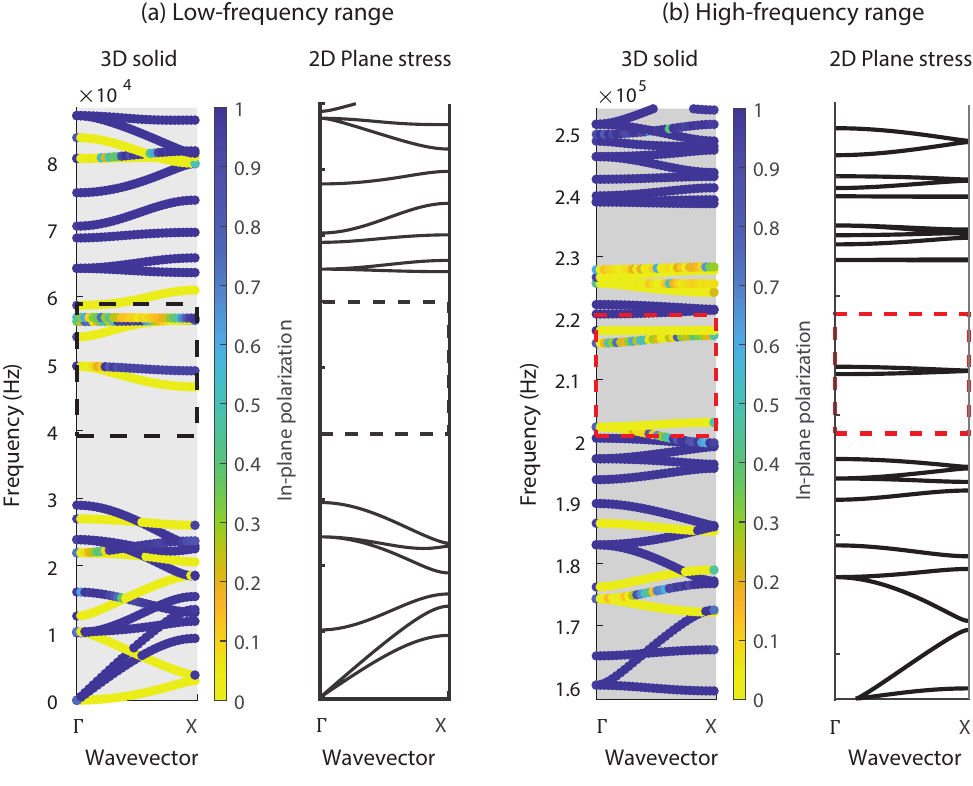}
    \caption{ unit-cell dispersion with the constitutive model of the wave-propagating medium chosen to be 3D elasticity (left), which better matches realistic plate structural behavior, and 2D plane stress (right), which is used for the template training dataset for computational efficiency. This is computed for both low- and high-frequency regimes in \textbf{(a)} and \textbf{(b)}, respectively. The color scale indicates the extent of wave in-plane polarization going from 0 (yellow) to 1 (blue).}
    \label{fig:Fig_09}
\end{figure}

To test the performance of the proposed hierarchical unit-cells in real finite engineering structures, a finite-structure case is considered. A steel periodic structure made out of a 8 by 8 unit-cell array of the top-left unit-cell in \autoref{fig:Fig_08} is taken as an example to illustrate the performance of the hierarchical unit-cell at multiple frequency ranges. The finite structure can be seen in \autoref{fig:Fig_11}, which shows the experimental setup used to validate our numerical computations.  The unit-cell side length is taken to be $L = 4$ cm and its thickness to be 0.635 cm. The sound speed in steel is computed as $c = \sqrt{E/\rho} = 5541.8$ m/s, where $E = 2\times10^{11} GPa$ and $\rho = 6.5 \times 10^3 $ kg/m$^3$. Thus, the frequency range of interest can be easily obtained by multiplying the normalized frequencies by the sound speed $c$ and dividing them by the unit-cell side length $L$. Since the unit-cell predictions are based on data of unit-cells modeled using 2D plane-stress, the dispersion relation is recomputed for a geometrically identical unit-cell modeled as a linear 3D solid. The dispersion curves of the 2D unit-cell modeled by plane stress and 3D elasticity are shown in \autoref{fig:Fig_09} in both the low- and high-frequency regimes (for coarse- and fine-scale bandgaps). It can be seen that both models yield remarkably similar results for in-plane waves. Some minor differences and frequency shifts can be noticed in the high-frequency regime, however the general trends are well preserved. This assures us of the soundness of our choice to use the computationally cheap 2D plane-stress constitutive model for our training dataset.
Thus, we have demonstrated the relevance of the hierarchical templates predictions to realistic plate structures.

\begin{figure}[htbp!]
\centering
\includegraphics[scale=0.6,trim={0cm 0cm 0cm 0cm},clip]{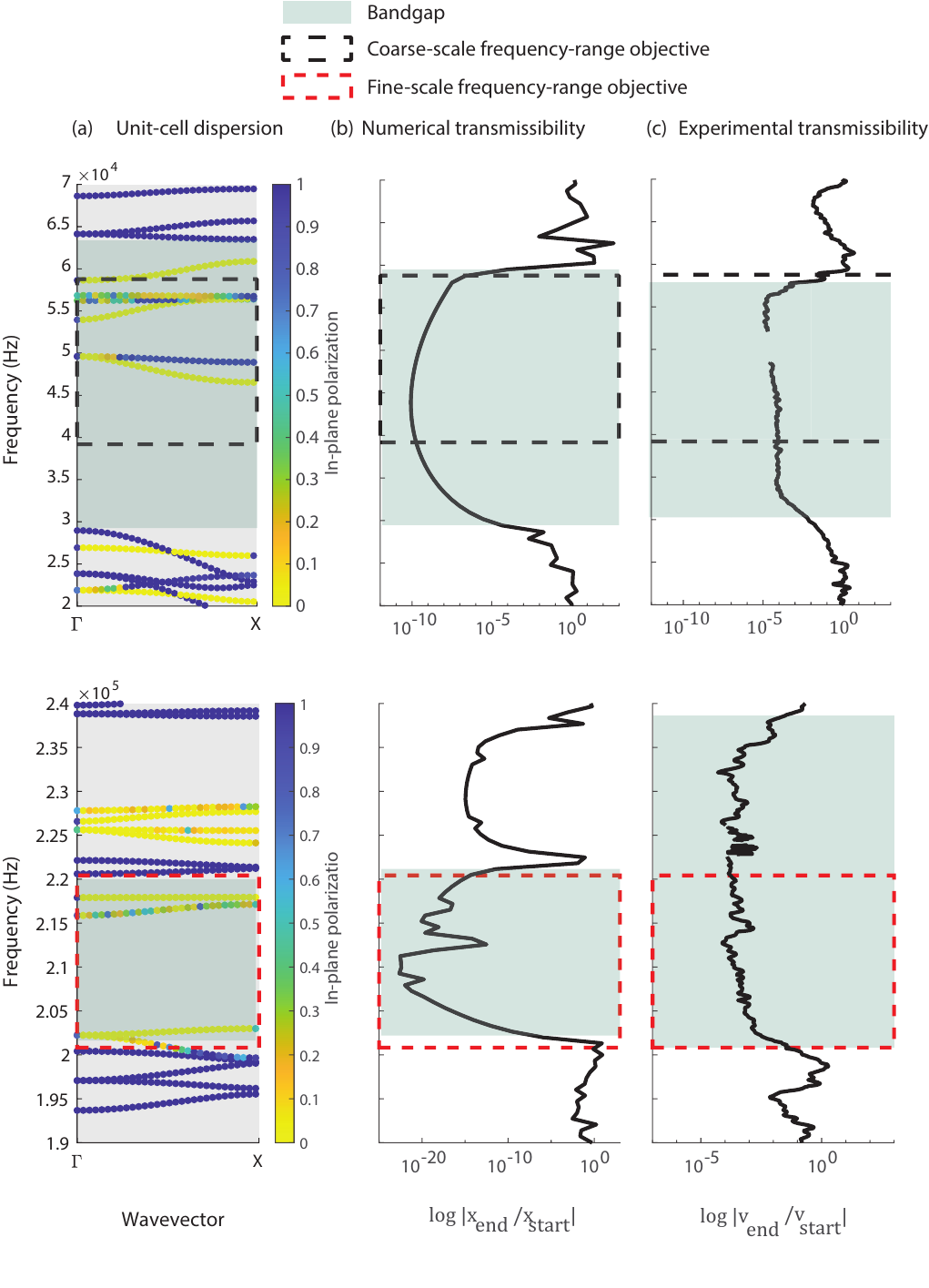}

    \caption{\textbf{(a)} Dispersion of the selected unit-cell modeled as a 3D solid with Bloch boundary conditions applied in the X- and Y-directions. The color scale indicates the extent of wave in-plane polarization going from 0 (yellow) to 1 (blue). Both the low-frequency (top) and high-frequency (bottom) ranges of interest are shown. \textbf{(b)} Numerical transmissibility of the structure modeled as a 3D elastic solid in both frequency ranges. \textbf{(c)} Experimental power transmissibility of the 3D printed structure in both frequency ranges.}
    \label{fig:Fig_10}
\end{figure}

As a final validation step for our method, the finite structure was tested and its response compared to our numerical predictions. The results are plotted in \autoref{fig:Fig_10}, in which the 3D elastic solid dispersion relation is shown next to the numerical and experimental transmissibility. The bandgap objective target regions are shown by the dashed boxes for both the low- and high-frequency ranges. A remarkable similarity in the transmission trends can be seen between both the numerical and experimental results. The bandgap regions also match well with the numerical and experimental response and it can be seen that the bandgap objectives are realized for in-plane waves (which are the target case) in both the low- and high-frequency bandgap objectives. More details surrounding these experiments can be found in \ref{sec:experimental_methods}.

\section{Methods}\label{sec11}

\subsection{Methods and algorithm workflow}
\label{sec:method}

We aim to obtain hierarchical templates for unit-cells that satisfy specified bandgap objectives in predefined frequency ranges for each hierarchical scale. 
Here, a two-level hierarchical template is used, though the algorithm can be expanded to include more levels. The bandgaps are specified to have a minimum width in order to be classified as bandgaps. 
Thus, our problem is a classification problem; the optimality of bandgap characteristics, like maximizing the bandgap width, is not the objective here.

In our hierarchical templates, we are considering solid-void phononic material unit-cells rather than the solid-solid unit-cells considered in prior work. This choice is made to both facilitate the fabrication process and to help make the structures lighter. Note that choosing void for one of the phases introduces additional challenges since many of the unit-cell configurations in the solid-solid design space are not valid for solid-void unit-cells, either because the solid parts are not connected, or because of other challenging fabrication issues. Thus, the unit-cell template method, reported in \cite{Chen2021-zx}, is extended here to include additional constraints that ensure the connectivity of the tessellated unit-cells and define a minimum finite feature size to prevent zero-dimensional singularity points that are not realistic to fabricate \cite{footnote3}. 

\begin{figure}[htbp!]
\centering
\includegraphics[scale=0.5,trim={0cm 0cm 0cm 0cm},clip]{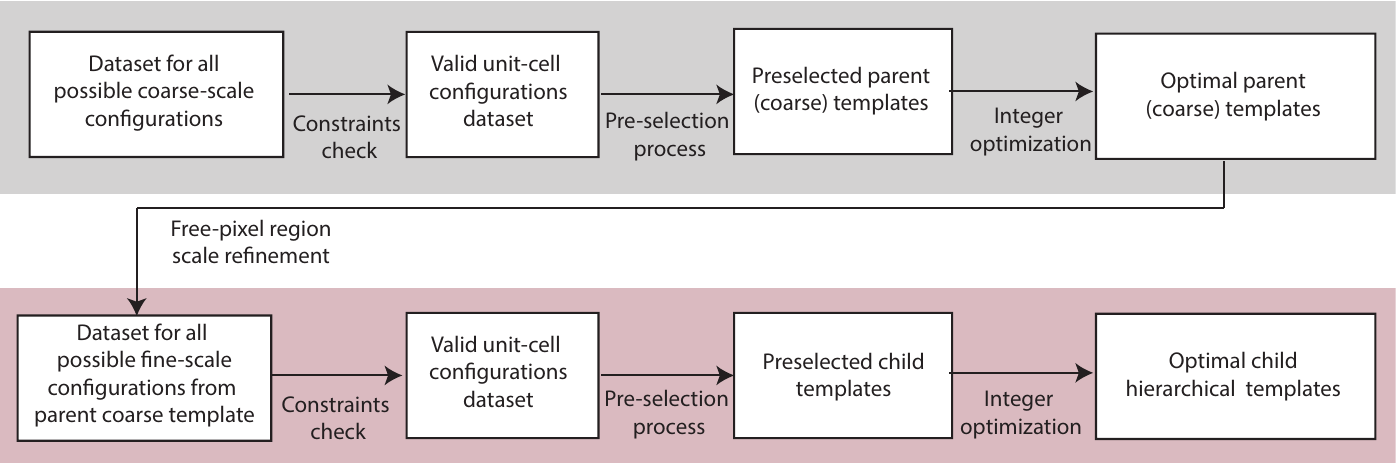}
    \caption{Outline for the hierarchical phononic-material unit-cell template method}
    \label{fig:Fig_03}
\end{figure}

As mentioned earlier and shown in \autoref{fig:Fig_02}, the process of obtaining the hierarchical unit-cell templates is performed consecutively starting at the coarse scale. The algorithm can be summarized in the following steps: 
\begin{itemize}
    \item We first generate the coarse-scale dataset to be used for obtaining the coarse-scale templates. The coarse-scale training data consists of all possible valid coarse-scale unit-cells, i.e., unit-cells satisfying connectivity and minimum finite feature size constraints, and their associated dispersion relations computed using the Finite Element method. This is typically computationally cheap as it involves only the coarse-scale design space. 
    \item Given the dataset, the template learning process contains 2 steps: (1) a pre-selection step that removes templates with low precision and support or that do not satisfy the manufacturability constraints, (2) an integer programming optimization that selects at most $s_{select}$ templates from the candidates to maximize the total support under the minimum precision constraint. 
    \item Using this template learning process, the coarse-scale ``parent'' templates are generated to open a bandgap in the first frequency range (left side of \autoref{fig:Fig_02} and coarse-scale bandgap in \autoref{fig:Fig_01} (a)).
    \item The parent templates are then refined to a finer scale and are used to generate the finer-scale dataset. Specifically, all solid and void pixels in the parent templates are kept to be the same, and each free pixel is split into $2\times 2$ finer-scale pixels. The refined free pixels are filled with all possible combinations of constituent materials and combined with the coarse-scale solid and void pixels to form the unit-cell designs in the finer-scale dataset. Again, Finite Element simulation is used to get the bandgap information, and only unit-cells constrained by connectivity and minimum feature size are included in the training data. Importantly, since the finer-scale dataset contains only unit-cells spanned by the refined free-pixel region, \textit{not the entire fine-scale unit-cells design space}, it greatly saves on the computational cost of acquiring data.
    \item The fine-scale templates or ``child templates'' (red grid on the right template in \autoref{fig:Fig_02})are optimized inside the fine-scale free-pixel regions to realize the fine-scale bandgap objective (red box in \autoref{fig:Fig_01}(a)). The optimization formulation is the same as that used for the coarse scale templates, i.e., a pre-selection step and integer programming optimization for selecting the optimal templates. The difference is that the design space is now a subset of the full space: it is now restricted to the fine-scale free-pixel regions since the other regions were fixed at the coarse scale.
    \item Hierarchical templates are thus obtained by attaching the fine-scale templates to the free-pixel regions of the coarse-scale templates. The coarse-scale features open the first bandgap, and the fine-scale features open the second bandgap (see \autoref{fig:Fig_02}).
\end{itemize}
The hierarchical template algorithm flow is also shown in \autoref{fig:Fig_03} and is detailed in \textbf{Section 1} of the supporting information. 

\subsection{Problem setup and definition}
\label{sec:setup}
As a demonstration, the design space is chosen here to be $k\times k$-square pixelated unit-cells with $p4mm$ symmetry (see \autoref{fig:Fig_02}). Due to unit-cell symmetry, unit-cells can be uniquely defined by only 1/8 of their area. The parameter $k$ is set to 10 in the coarse-scale design space and $k$ is 20 in the fine-scale design space. For the $10\times 10$-pixel coarse design space, the unit-cells can be uniquely defined by 15 pixels, resulting in $2^{15}$ unit-cell configurations. A plane-stress solid-mechanics model is used to model the behavior of plate-type media. The dataset containing all coarse-grid ($10\times 10$) unit-cells and their dispersion relations is created using an in-house developed  finite-element code. More details on the finite-element implementation can be found in \textbf{Section 2} in the supporting information. 

The utilized target frequency ranges and constraints in the optimization process for the hierarchical templates are shown in \autoref{tab:TempOptParams}. As an example, two pairs of normalized frequency ranges are chosen for the coarse- and fine-scale bandgap target range, \{(0.28, 0.42),(0.57, 0.71)\} and \{(0.28, 0.42),(1.45, 1.59)\}. The objective is to obtain a bandgap covering a percentage of each target range. The desired percentage coverage is further specified to be entirely contained within a single bandgap, not as a sum of multiple ones. Choices for the minimum optimal precision and the maximum optimal template set size are also shown in \autoref{tab:TempOptParams}. 

\begin{table}[htbp!]
\centering
\begin{tabular}{p{2cm}|p{1.5cm}|p{1.5cm}|p{3cm}|p{2.7cm}}
\hline
Optimization process  & Minimum precision constraint & Maximum optimal template-set size & Target normalized-frequency ($f L c^{-1}$) range & Minimum bandgap width (target-range coverage $\%$) \\ \hline
Parent template & 99\% & 9 & (0.28, 0.42) & 60\\
Hierarchical template & 99\% & 16 & \{(0.28, 0.42),(0.57, 0.71)\} & \{60, 35\}\\
Hierarchical template & 99\% & 16 & \{(0.28, 0.42),(1.45, 1.59)\} & \{60, 35\}\\
\hline
\end{tabular}

\caption{Constraints of the hierarchical templates optimization process}
\label{tab:TempOptParams}
\end{table}

\subsection{Experimental methods}
\label{sec:experimental_methods}

\begin{figure}
    \centering
    \includegraphics[width = 0.45\linewidth, angle = 0]{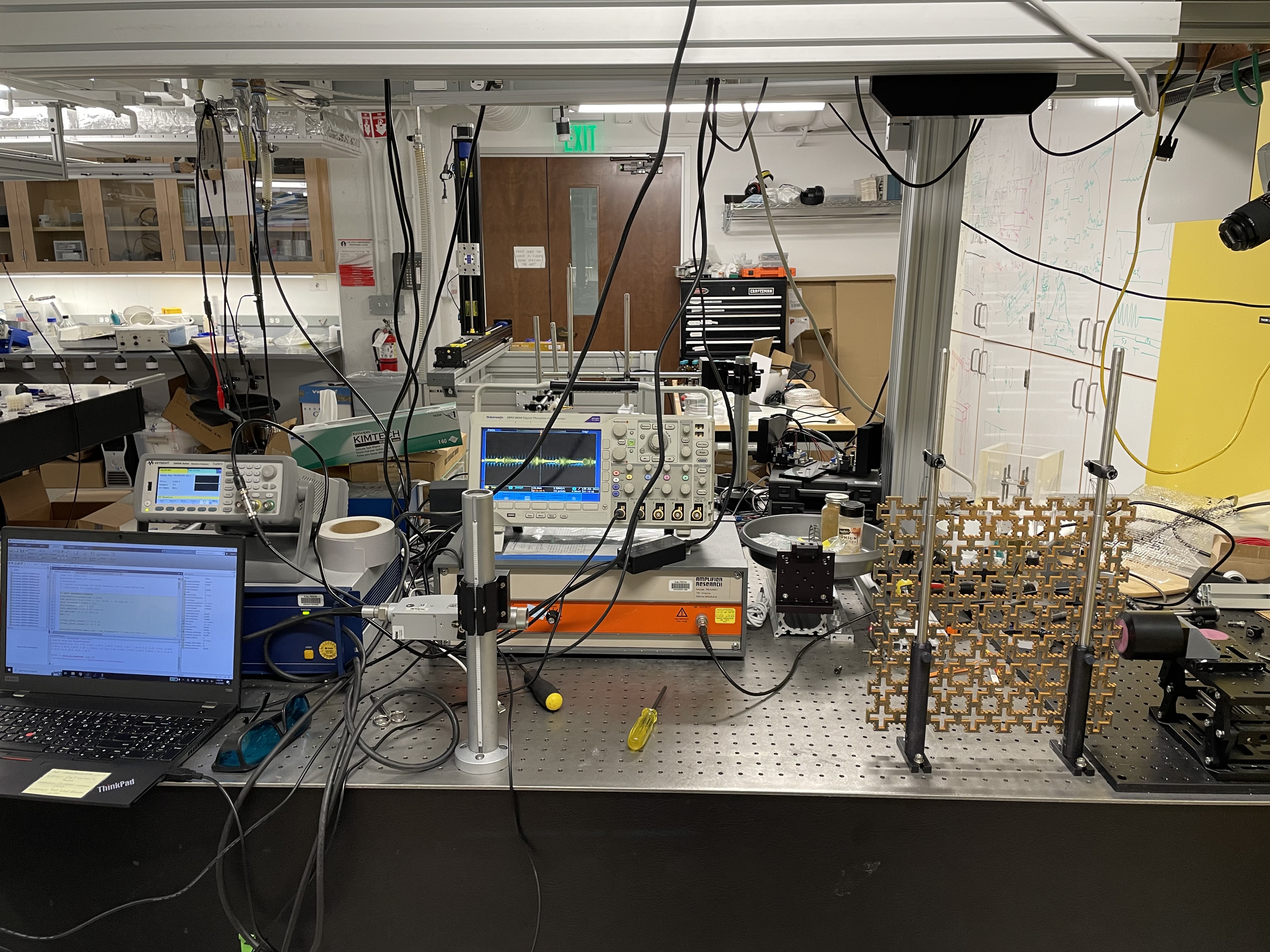}
    \includegraphics[width = 0.45\linewidth, angle = 0]{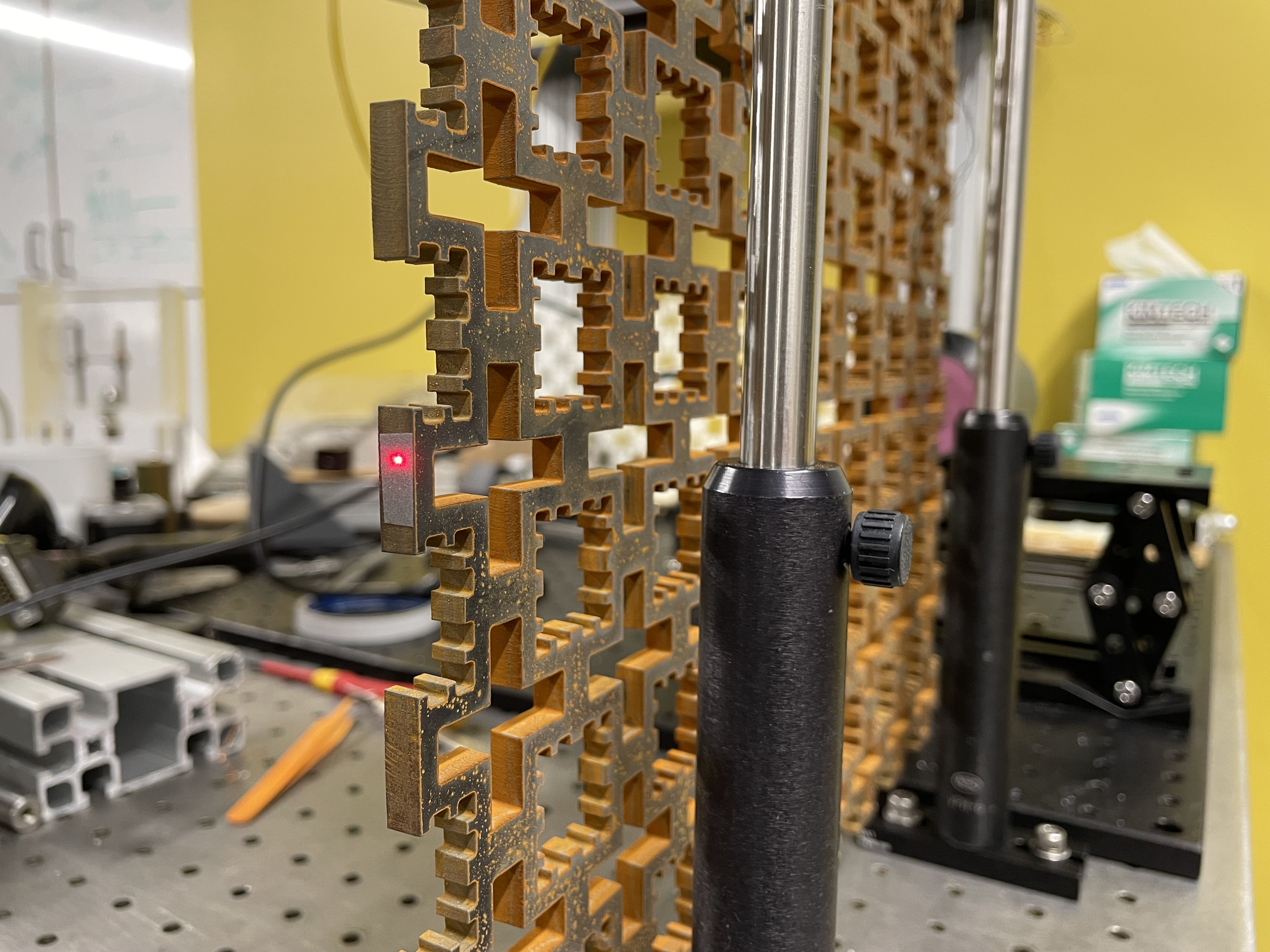}
    \caption{The experimental setup. On the left, a view of how the equipment and sample are assembled. On the right, a view of the LDV measurement point.}
    \label{fig:Fig_11}
\end{figure}

Experiments were carried out to validate the simulations by verifying the existence of bandgaps.
An 8 by 8 lattice of 4cm square unit-cells was waterjet cut from 1/4 inch thick steel.
The frequency response function of the structure was measured.
A piezoelectric transducer was used to provide the excitation, and a laser Doppler vibrometer (LDV) was used to measure the input signal and output signal.
An image of the setup is shown in Fig. \ref{fig:Fig_11}.

To measure the frequency response function, a chirp signal was sent through the structure. 
A chirp is a sinusoid signal with a continuously varying frequency, and is an excitation commonly used for frequency response measurements.
To post-process the results, we use spectrogram analysis.
A spectrogram is a time-varying power spectrum; it shows how the power spectrum of a signal changes with time.

The structure was suspended at two points using a thin cable to approximate free-floating boundary conditions.
At one end, excitation was provided by a piezo-based ultrasonic transducer (Panametrics Videoscan V1012), used in tandem with an amplifier (Amplifier Research Model 75A250).
Input signals were generated by a Keysight 33500B Series waveform generator.
The waveform was custom generated on the laptop using Matlab.
To measure and store signals, we used a Laser Doppler Vibrometer (LDV) (Polytec CLV-2534), connected to a Tektronix DPO 3014 oscilloscope.
The LDV converted observed velocities into voltages to produce an electrical signal.
This signal was then read and stored by the oscilloscope, and subsequently sent to the computer.

\section{Discussion}\label{sec13}
Hierarchical phononic-material templates provide a a new approach to discover multiscale hierarchical patterns in unit-cells that satisfy bandgap design objectives for each length scale. The templates are obtained by extending the unit-cell template method, an interpretable machine learning technique, to deal with multiscale grids and multiple bandgap design objectives. The advantages of our approach are as follows: 
\begin{itemize}
    \item This approach is robust to changes introduced on the fine-scale, while keeping the coarse scale objectives satisfied. This is true even when the coarse and fine hierarchical length scales are close. This robustness is consistently demonstrated and its precision is quantified.  Thus, a scale-separation effect emerges as a result of this robustness.
    \item  The scale-separation effect enables us to obtain templates successively at each length scale starting from the coarse scale, offering a tremendous reduction of computational cost of unit-cell design. This is in contrast to the currently used computationally expensive approaches that use the fine-scale grid design space directly. Using the fine-scale grid has been typical when the hierarchical length scales are not well separated.
    \item The computational savings of our approach allows us to discover hierarchical patterns in previously unexplored areas of the design space that depart from the commonly-used self-similar or fractal patterns. Moreover, many hierarchical unit-cells can be generated from a single hierarchical template by varying the constituents of the free-pixel regions, adding more flexibility and insight into the relevant patterns for generating materials with specific properties. 
    \item The presented results and learned patterns here apply to bandgap objectives in solid/void unit-cells with any solid material choice that can be modeled as linear elastic. The approach can be generalized to learn different wave-dispersion design objectives, or to be used with unit-cells made of multiple materials.
\end{itemize}
We demonstrated numerically and experimentally how the hierarchical phononic templates can be successfully utilized to design 3D finite structures that satisfy predefined bandgap objectives at different frequency ranges. Thus, the hierarchical phononic material templates offer a new design paradigm that promises to make the design process more flexible, computationally cheaper, with the added value of highlighting global patterns for achieving predefined design objectives.   

\bmhead{Supplementary information}

Supporting Information is available online or from the corresponding author.

\bmhead{Data availability}

The data used to generated these results can be requested from the corresponding author. 

\backmatter

\bmhead{Acknowledgements}

Funding: We acknowledge support for this work from the Department of Energy under grant DE-SC0021358, and the National Science Foundation under grants DGE-2022040 and OAC-1835782.

\bmhead{Conflicts of interest}

The authors declare no competing financial or non-financial interests.

\bmhead{Authors' contributions}
Mary Bastawrous and Zhi Chen conceptualized the project, led the primary research activities for the project, and led the writing and revising of the manuscript.
Alex Ogren completed the experimental validation and contributed to writing and revising the manuscript.
Cynthia Rudin and Cate Brinson participated in guiding the project and revising the manuscript.

\bibliography{HMrefs}

\begin{thebibliography}{10}
\bibitem{footnote3}
\bibinfo{title}{Note that the singularity point are demonstrated later in
  \autoref{fig:Fig_06}.}
\end{thebibliography}

\end{document}